\documentclass[twocolumn,showpacs,preprintnumbers,amsmath,amssymb]{revtex4}

\usepackage{graphicx}
\usepackage{dcolumn}
\usepackage{bm}
\usepackage{booktabs}
\usepackage{multirow}
\usepackage{amssymb}

\begin{document}


\title{Estimation of the spin polarization for Heusler-compound thin films by means of nonlocal spin-valve measurements: \\Comparison of Co$_{2}$FeSi and Fe$_{3}$Si}

\author{K. Hamaya,$^{1,2}$\footnote{E-mail: hamaya@ed.kyushu-u.ac.jp} N. Hashimoto,$^{1}$ S. Oki,$^{1}$ S. Yamada,$^{1}$ M. Miyao,$^{1,3}$ and T. Kimura$^{4,3}$\footnote{E-mail: kimura@ifrc.kyushu-u.ac.jp}}
\affiliation{%
$^{1}$Department of Electronics, Kyushu University, 744 Motooka, Fukuoka 819-0395, Japan}%
\affiliation{%
$^{2}$PRESTO, Japan Science and Technology Agency, Sanbancho, Tokyo 102-0075, Japan}%
\affiliation{%
$^{3}$CREST, Japan Science and Technology Agency, Sanbancho, Tokyo 102-0075, Japan}%
\affiliation{
$^{4}$INAMORI Frontier Research Center, Kyushu University, 744 Motooka, Fukuoka 819-0395, Japan}%
\date{\today}

\begin{abstract}
We study room-temperature generation and detection of pure spin currents using lateral spin-valve devices with Heusler-compound electrodes, Co$_{2}$FeSi (CFS) or Fe$_{3}$Si (FS). The magnitude of the nonlocal spin-valve (NLSV) signals is seriously affected by the dispersion of the resistivity peculiarly observed in the low-temperature grown Heusler compounds with ordered structures. From the analysis based on the one-dimensional spin diffusion model, we find that the spin polarization monotonically increases with decreasing the resistivity, which depends on the structural ordering, for both CFS and FS electrodes, and verify that CFS has relatively large spin polarization compared with FS.  
 \end{abstract}

\maketitle
In the field of spintronics, the evaluation of the spin polarization of ferromagnetic materials is essential for understanding the general physics on materials and spin-related transport properties in device structures. To realize high-performance device applications, the use of highly spin-polarized ferromagnets is required. The Co-based Heusler compounds, expected to be a half-metallic ferromagnet (HMF),\cite{Galanakis,Felser,Wurmehl,Inomata,Katsnelson} are promising to obtain the huge magnetoresistance effects in the vertical device structures\cite{Inomata,Marukame,Sakuraba1,Hono} and to demonstrate electrical spin injection into semiconductors.\cite{Herfort1} 

The spin polarizations of these Co-based Heusler compounds have so far been evaluated from the direct measurements with the point contact Andreev reflection (PCAR) techniue,\cite{Ritchie} the Valet-Fert model with the current-perpendicular-to-plane giant magnetoresistance (CPP-GMR) effect,\cite{Sakuraba2,Nakatani} and the Julliere's model with the tunneling magnetoresistance (TMR) effect.\cite{Sukegawa,Yamamoto} As a result, relatively high values have been reported. However, it is technically difficult to obtain the precise spin polarization at room temperature for these Heusler compounds. First of all, the PCAR method is limited only at low temperatures unfortunately though it provides the comparatively precise estimation of the spin polarization. For the Valet-Fert model with CPP-GMR, one can obtain only the limited information about a series resistance consisting of the ferromagnetic electrodes, nonmagnetic spacer, and the interfaces.\cite{Sakuraba2,Nakatani} For the Julliere's model with TMR, the tunnel barrier used in the most of the magnetic tunnel junctions is composed of the crystalline MgO because of the sequential epitaxial growth for the top Heusler-compound electrode.\cite{Sukegawa,Yamamoto} In this case, one may overestimate the value of the spin polarization of the Heusler-compound electrodes used if the spin-filter effect of the MgO barrier contributes predominantly the enhancement in the TMR effect.\cite{Yuasa,Parkin} Furthermore, for the above two vertical structures, the bottom and top electrodes are fabricated in different conditions whereas the spin polarization of the Heusler compounds can sensitively change with the degree of the structural ordering.\cite{Inomata,Sakuraba2,Sukegawa} 
\begin{figure}[b]
\includegraphics[width=5cm]{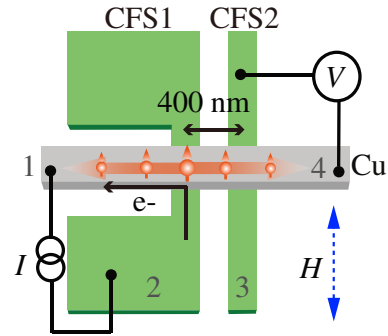}
\caption{(Color online) Schematic of a nonlocal spin valve measurement. Spin-polarized electrons are injected from contact 2, and electron charges are extracted from contact 1. A nonlocal voltage is measured between contact 3 and contact 4.}
\end{figure}

On the other hand, multi-terminal laterally configured structures can provide us many information on spin-related phenomena because of the flexible probe configurations. In particular, the nonlocal spin valve (NLSV) measurements enable us to detect the pure spin current generated in nonmagnets in mesoscopic devices.\cite{Jedema,Jedema1,Takahashi,Kimura,Valenzuela1,Ji,Lou} In addition, one can separately measure the resistances of the ferromagnetic electrodes, the nonmagnetic wires and the ferromagnet-nonmagnet interfaces. Furthermore, the used spin injector and detector consist of the same ferromagnetic layer grown on the same substrate in the lateral device structures. Therefore, the use of the lateral structures combined with the NLSV measurements is effective to evaluate the spin polarization for the Co-based Heusler-compound thin films. Recently, we have observed giant spin signals at room temperature using lateral spin-valve devices with Co$_{2}$FeSi (CFS) electrodes.\cite{KimuraHamaya} Also, Bridoux {\it et al}. demonstrated the similar features at 77 K using the incidentally formed Co$_{2}$FeAl electrodes.\cite{Bridoux} These two results indicate the great potential of Co-based Heusler-compounds with half metallicity for generating pure spin currents. 

In this paper, we study room-temperature generation and detection of pure spin currents in LSV devices with Co$_{2}$FeSi (CFS) electrodes. For comparison and verification, one of the Heusler compounds, Fe$_{3}$Si (FS), is also explored. Here our low-temperature grown CFS and FS films include $L2_{1}$ and $D0_{3}$ ordered structures,\cite{Yamada,Hamaya} respectively. The magnitude of the nonlocal spin-valve (NLSV) signals is seriously affected by the dispersion of the resistivity for both Heusler-compound electrodes. From the analysis based on the one-dimensional spin diffusion model, we find that the spin polarization monotonically increases with decreasing the resistivity for both CFS and FS electrodes, and verify that CFS has relatively large spin polarization compared with FS.
\begin{figure}
\includegraphics[width=8cm]{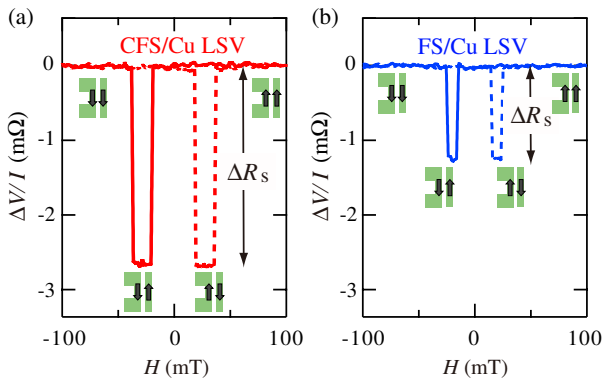}
\caption{(Color online) Room-temperature nonlocal spin-valve signals for (a) CFS/Cu LSV and (b) FS/Cu LSV. The signal varies according to the relative magnetization orientation of two wire-shaped electrodes, as schematically illustrated.}
\end{figure}  

As spin injector and detector materials, 25-nm-thick Co$_{2}$FeSi (CFS) and Fe$_\text{3}$Si (FS) films were grown on Si(111) by LT-MBE.\cite{Yamada,Hamaya} Here FS is expected not to be half-matallic.\cite{Bansil} The epitaxial CFS and FS films formed were characterized by means of cross-sectional transmission electron microscopy (TEM), nanobeam electron diffraction (ED), and $^{57}$Fe conversion electron M\"ossbauer spectroscopy. From these detailed characterizations, we have observed $L2_{1}$ and $D$0$_\text{3}$ ordered structures in the CFS and FS layers, respectively, where the degree of the local structural ordering is more than 60 \%.\cite{Yamada,Hamaya2} Using conventional electron-beam lithography and an Ar ion milling technique, we patterned wire-shaped CFS or FS spin injector and detector with $\sim$250 nm in width. One CFS or FS wire is connected with two rectangular pads to facilitate domain wall nucleation. The Cu wire with $\sim$250 nm in width and bonding pads were fabricated by a conventional lift-off technique. Prior to the Cu deposition, the surface of the CFS and FS wires was well cleaned by using the Ar ion milling with a low accelerating voltage, resulting in low resistive ohmic interfaces ($<$ 0.1 ${\rm f}\Omega {\rm m}^2$). Using the two different wire shapes, we can control the magnetization configuration by adjusting external magnetic fields ($H$), where $H$ is applied along the wires. Nonlocal and local spin valve measurements were carried out by a conventional current-bias lock-in technique ($\sim 200$ Hz, $\sim 100 \mu$A). A pure spin current generated by the nonlocal spin injection from CFS1 can be detected by CFS2 after the propagation of 400-nm distance in the Cu strip, as schematically shown in Fig. 1. All the nonlocal spin-valve measurements were performed in the cross configuration in order to maintain the validity of the analysis based on one dimensional model discussed below.\cite{Hamrle} 
\begin{figure}
\includegraphics[width=8cm]{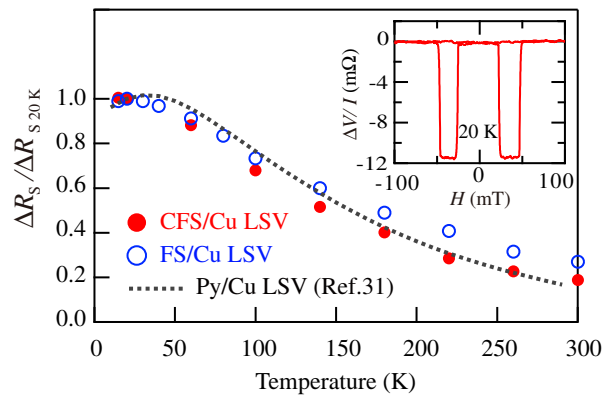}
\caption{(Color online) Temperature dependence of $\Delta R_{\rm S}$ for the CFS/Cu (red solid circles) and FS/Cu (blue solid circles) LSVs, normalized by $\Delta R_{\rm S}$ at 20 K. The black dashed curve is the data for Py/Cu LSV reported in Ref. \cite{Kimura2}. The inset shows a nonlocal spin-valve effect of the CFS/Cu LSV at $T = 20$ K.}
\end{figure}

Figure 2(a) shows a nonlocal magnetoresistance of a CFS/Cu LSV, measured at room temperature (RT). For comparison, that of an FS/Cu LSV is also shown in Fig. 2(b). Here the size of the CFS/Cu junction ($\sim$ 250 $\times$ 250 nm$^{2}$) is almost the same as that of the FS/Cu junction. By sweeping $H$, the relative magnetization orientation of two wire-shaped electrodes was controlled, as depicted in the inset illustrations. For both LSVs we observe clear NLSV signals. It should be noted that a giant spin signal ($\Delta R_{\rm S}$) of 2.6 m$\Omega$ is seen for the CFS/Cu LSV [Fig. 2(a)], which is approximately two times as large as that for the FS/Cu LSV. The $\Delta R_{\rm S}$ value of 2.6 m$\Omega$ was ten times larger than that for the Py/Cu LSVs with the same size and low resistive ohmic interfaces. We also obtained the local spin valve signal of 4.9 m$\Omega$ at RT for the same CFS/Cu LSV. The value of 4.9 m$\Omega$ is almost twice of the non-local $\Delta R_{\rm S}$, in reasonable agreement with the previous reports.\cite{Jedema,Jedema1,Kimura,Takahashi} We have already confirmed that $\Delta R_{\rm S}$ measured in the half configuration is 15 \% larger than that in the cross configuration. The probe-configuration dependence can be quantitatively explained by the difference in the effective distance from the injector to the detector, similar to the previous Py/Cu LSVs.\cite{Hamrle} Almost the same features were also seen in the FS/Cu LSV. From these facts, one dimensional spin diffusion model well describes the spin transport properties in the present CFS/Cu and FS/Cu LSVs.

Next, the temperature dependence of $\Delta R_{\rm S}$ for the representative CFS/Cu and FS/Cu LSVs is examined in Fig. 3, where $\Delta R_{\rm S}$ is normalized by $\Delta R_{\rm S}$ at 20 K ($\Delta R_{\rm S}$/$\Delta R_{\rm S 20K}$). The temperature evolution for the CFS/Cu and FS/Cu LSVs is almost equal to that for the Py/Cu LSV reported in Ref.\cite{Kimura2}. Thus we regard this feature as a consequence of an enhancement in the spin-flip scattering in the Cu wire.\cite{Kimura2} Judging from the results shown in Figs. 2 and 3, we can recognize that the LSVs presented here can be treated as conventional ohmic LSVs. The inset shows a NLSV effect of the CFS/Cu LSV at $T = 20$ K. Surprisingly, the $\Delta R_{\rm S}$ exceeds 10 m$\Omega$ at 20 K even for the device with ohmic interfaces and relatively large lateral dimensions.
 
Hereafter we focus on the value of $\Delta R_{\rm S}$. For our CFS and FS LSVs fabricated, we often observed the dispersed $\Delta R_{\rm S}$ values. Since it is well known that the area of the junctions affects the value of $\Delta R_{\rm S}$ in NL measurements,\cite{Kimura} we introduce the $A$ value defined as ($S_{\rm inj}$$S_{\rm det}$/$S_{\rm N}$), where $S_{\rm inj}$, $S_{\rm det}$, and $S_{\rm N}$ are the areas of the junctions in spin injector, spin detector, and the cross section of the Cu strip, respectively. In our previous works,\cite{KimuraHamaya} we simplified the spin signals with considering $A$ on the basis of a one-dimensional spin diffusion model.\cite{VFmodel,Takahashi} 
\begin{equation}
\Delta R_{\rm S} A \approx 
\frac{ \left( \frac{ P_F}{(1-P_F^2)} \rho_{\rm F} \lambda_{\rm F} 
+ \frac{P_I}{(1-P_I^2)} RA_{\rm F/N} \right)^2 }
{\rho_{\rm N} \lambda_{\rm N} \sinh \left( {d}/{\lambda_{\rm N}} \right)},
\end{equation}
where $P_{\rm F}$ and $P_{\rm I}$ are the bulk and interface spin polarizations of the ferromagnetic electrode, respectively. $\rho_{\rm F}$ and $\lambda_{\rm F}$ are the resistivity and the spin diffusion length of the ferromagnetic electrode, and $\rho_{\rm N}$ and $\lambda_{\rm N}$ are those for the nonmagnetic wire and $d$ is the separation distance between the injector and detector. In this study, $d$ is a constant value of 400 nm. $RA_{\rm F/N}$ is the interface resistance between ferromagnet and nonmagnet, but in this study we can ignore this term because of $RA_{\rm F/N}$ $<$ 0.1 ${\rm f}\Omega {\rm m}^2$. The detailed procedure of the introduction of Eq. (1) is reported in elsewhere.\cite{KimuraHamaya} If we consider $\Delta R_{\rm S}$$A$ in Eq. (1), the influence of uneven $A$ on the spin signal can be normalized. Thus, $\Delta R_{\rm S} A$ allows us to fairly evaluate the generation efficiency of the pure spin current for various junctions. We want to call $\Delta R_{\rm S} A$ the resistance change area product for the NLSV signal. To take into account $\Delta R_{\rm S} A$, we measured the accurate $A$ values of the CFS/Cu or FS/Cu junctions by directly observing SEM images. After the normalization of the geometrical factor, we still observed the dispersion of $\Delta R_{\rm S} A$. Thus the dispersed $\Delta R_{\rm S}$ values for our LSVs originate mainly from other than geometrical factors. 

We also note that $\rho_{\rm F}$ is strongly related to $\Delta R_{\rm S}$$A$ in Eq. (1). In general, it is well known that there is the fluctuation of the long-range atomic ordering in the epitaxial Heusler-compound films grown by LT-MBE. In fact, the large changes in $\rho_{\rm F}$, depending on the atomic ordering, have already been reported for $L2_{1}$-ordered CFS/GaAs\cite{Hashimoto} and $D$0$_\text{3}$-ordered FS/GaAs\cite{Herfort} in the previous works. The magnitude of $\rho_{\rm F}$ for our CFS and FS films is almost the same orders as the previous works.\cite{Hashimoto,Herfort,Schneider} Considering these characteristic trends of epitaxial Heusler-compound films grown by LT-MBE, we summarize $\Delta R_{\rm S}$$A$ as a function of $\rho_{\rm F}$ for CFS/Cu and FS/Cu LSVs in Fig. 4(a). Since there are relatively wide dispersions of $\rho_{\rm F}$ for both CFS and FS, this figure clearly shows that the dispersed $\Delta R_{\rm S}$$A$ for LSVs is markedly associated with the dispersion of $\rho_{\rm F}$ for spin injector and detector. 
\begin{figure}
\includegraphics[width=8.5cm]{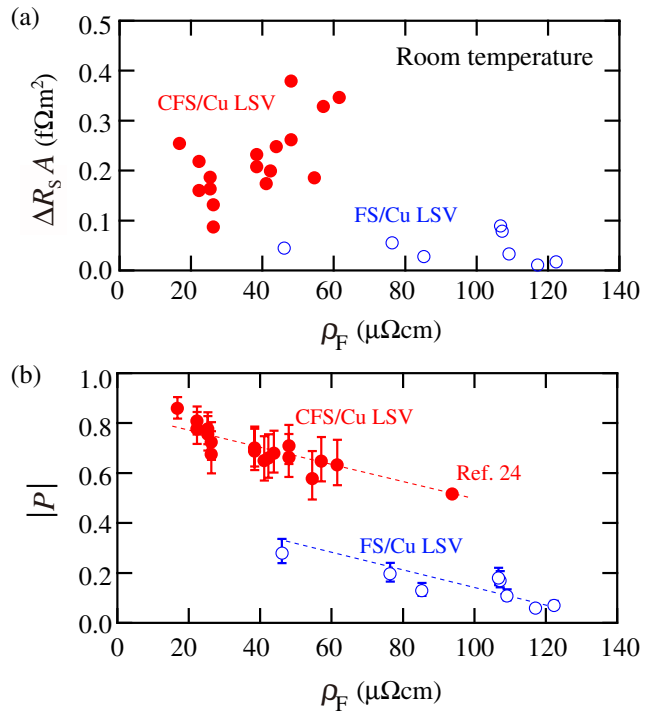}
\caption{(Color online) (a) $\Delta R_{\rm S}$$A$ as a function of $\rho_{\rm F}$ at room temperature for CFS/Cu and FS/Cu LSVs. (b) The trend of spin polarization ($P$) estimated from a simple one-dimensional spin diffusion model for CFS/Cu and FS/Cu LSVs. The used $\lambda_{\rm F}$ values are 3 $\pm$ 1 nm ($\lambda_{\rm CFS}$) and 5 $\pm$ 1 nm ($\lambda_{\rm FS}$), respectively. }
\end{figure}

Using Eq.(1), we can roughly estimate $P$ from the obtained $\Delta R_{\rm S}$$A$ with an assumption of $\lambda_{\rm F}$. Even for the high performance Co-based Heusler compounds,\cite{Nakatani} $\lambda_{\rm F}$ value of 3 $\pm$ 1 nm was reported, which is the same orders of magnitude for conventional ferromagnetic alloys.\cite{Starikov} Thus, we can assumedly regard $\lambda_{\rm F}$ for our CFS as  3 $\pm$ 1 nm at room temperature ($\lambda_{\rm CFS}$).\cite{Nakatani} Here for Fe-based Heusler compounds there is almost no information on $\lambda_{\rm F}$. Since $\lambda_{\rm F}$ for Fe films was reported as 8 nm at 4.2 K,\cite{Bass} $\lambda_{\rm F}$ for Fe-based alloys is naively expected to be shorter than 8 nm at 4.2 K. At room temperature, further short $\lambda_{\rm F}$ for Fe-based alloys can be expected. Thus, $\lambda_{\rm F} =$ 8 nm can be considered to be the upper limit of $\lambda_{\rm F}$ for our FS. On the other hand, for Py (one of the alloys including Fe), $\lambda_{\rm F}$ was reported to be 2 nm at room temperature.\cite{Kimura,Dubois} This can be considered to be the lower limit. Considering the assumable upper and lower limits of $\lambda_{\rm F}$ for our FS ($\lambda_{\rm FS}$), we should realistically regard $\lambda_{\rm FS}$ as 5 $\pm$ 1 nm at room temperature. The estimated $P$ as a function of $\rho_{\rm F}$ is shown in Fig. 4(b), where $\rho_{\rm N}$ and $\lambda_{\rm N}$ are 2.5 $\mu$$\Omega$cm and 500 nm,\cite{Kimura,KimuraHamaya} respectively. The $P$ values for CFS/Cu LSVs are larger than those for FS/Cu ones. In this figure, we can also see nearly monotonic enhancement in $P$ with decreasing $\rho_{\rm F}$. This fact means that highly ordered Heusler-compound electrodes have large $P$, consistent with general interpretations.\cite{Blum,Hashimoto,Herfort,Schneider} Note that this correlation between $P$ and $\rho_{\rm F}$ can also reproduce our previous result ($P \sim$ 0.56 for $\rho_{\rm CFS} \sim$ 90 $\mu$$\Omega$cm).\cite{KimuraHamaya} For these reasons, this analysis based on the NLSV measurements and the one-dimensional spin diffusion model is a universal method for fairly evaluating the spin polarization of Heusler-compound thin films. Since the large $\rho_{\rm F}$ value is effective to enhance $\Delta R_{\rm S}$$A$,\cite{Kimura} one should not simply regard the large $\Delta R_{\rm S}$$A$ as a consequence of the large $P$ in Heusler compounds, as shown in Figs. 4(a) and 4(b).

We finally comment on the validity of the analysis described above. In Fig. 4(a) all the $\Delta R_{\rm S}$$A$ values for CFS/Cu LSVs fabricated are larger than those for FS/Cu LSVs even if we take into account the dispersion of $\rho_{\rm F}$. In comparison with FS/Cu LSVs (higher $\rho_{\rm F}$), CFS/Cu LSVs have relatively large $\Delta R_{\rm S}$$A$ values despite the generally lower $\rho_{\rm F}$. This indicates that the relatively large $\Delta R_{\rm S}A$ observed in CFS/Cu LSVs are not arising from the large $\rho_{\rm F}$ of CFS. The large $P$ or long $\lambda_{\rm F}$ for CFS should be considered as an origin of the large $\Delta R_{\rm S}$$A$ for CFS/Cu LSVs. Recently, we obtained the relatively large $P$ value ($P =$ 0.59) for one of our CFS films,\cite{Yamada2} measured by the PCAR method. This value is larger than that of FS films ($P =$ 0.45) reported by Ionescu {\it et al}.\cite{Ionescu} On the other hand, $\lambda_{\rm F}$ for CFS(Al) can be predicted to be relatively short ($ \lambda_{\rm F} =$ 3 $\pm$ 1 nm),\cite{Nakatani} almost the same as conventional ferromagnets.\cite{Bass,Starikov} Considering these actual data, we infer that the large $P$ of our CFS is a main origin of the relatively large $\Delta R_{\rm S}$$A$ for CFS/Cu LSVs compared with FS/Cu LSVs. 

In summary, we have studied generation and detection of pure spin currents at room temperature using lateral spin-valve devices with Heusler-compound electrodes, Co$_{2}$FeSi (CFS) or Fe$_{3}$Si (FS). From the analysis based on the one-dimensional spin diffusion model, we obtained monotonic correlation between spin polarization and resistivity for both CFS and FS LSVs. We verified that CFS has relatively large spin polarization compared with FS by means of NLSV measurements.

This work was partially supported by CREST-JST, and TEPCO Memorial Foundation, and Grant-in-Aid for challenging Exploratory Research from JSPS. 


\end{document}